\documentclass[a4paper,12pt]{article}
\usepackage{epsfig}
\begin{document}

\begin{center}

{\large\bf A new scheme for heavy nuclei: proxy-SU(3)} \bigskip
 
 \footnotesize

D. Bonatsos$^1$, R. F. Casten$^{2,3}$, A. Martinou$^1$, I. E. Assimakis$^1$, N. Minkov$^4$, S. Sarantopoulou$^1$, 

 R. B. Cakirli$^5$, and K. Blaum$^6$

\medskip

{\it 
$^1$ Institute of Nuclear and Particle Physics, National Center for Scientific Research  ``Demokritos'', Athens, Greece 

$^2$ Wright Laboratory, Yale University, New Haven, Connecticut 06520, USA  

$^3$ Facility for Rare Isotope Beams, 640 South Shaw Lane, Michigan State University, East Lansing, 

MI 48824 USA  

$^4$ Institute of Nuclear Research and Nuclear Energy, Bulgarian Academy of Sciences, 

72 Tzarigrad Road, 1784 Sofia, Bulgaria

$^5$ Department of Physics, University of Istanbul, Istanbul, Turkey  

$^6$ Max-Planck-Institut f\"{u}r Kernphysik, Saupfercheckweg 1, D-69117 Heidelberg, Germany } 

\end{center}

\rule{16.5cm}{0.3mm}

\bigskip

\footnotesize
{\bf Abstract} \hskip 5mm  The SU(3) symmetry realized by J. P. Elliott in the sd nuclear shell is destroyed in heavier shells 
by the strong spin-orbit interaction. However, the SU(3) symmetry has been used 
for the description of heavy nuclei in terms of bosons in the framework of the Interacting Boson 
Approximation, as well as in terms of fermions using the pseudo-SU(3) approximation. We introduce a new fermionic approximation, called the proxy-SU(3), and we discuss how some of its novel predictions come out as 
a consequence of the short range of the nucleon-nucleon interaction and the Pauli principle. 

\bigskip


 \rule{16.5cm}{0.3mm}
\pagestyle{empty}

\normalsize

\bigskip\noindent
{\bf INTRODUCTION}

\medskip\noindent

A new algebraic approach to heavy deformed nuclei, called the proxy-SU(3) scheme, has been introduced recently \cite{proxy1,proxy2}. Proxy-SU(3) is based on fermionic symmetries. A microscopic justification in terms of a Nilsson model calculation has been discussed in Ref. \cite{proxy1}. Parameter-free predictions for the deformation parameters $\beta$ and $\gamma$ of the collective model for even rare earth nuclei have been given in Ref. \cite{proxy2} and successfully compared to mean field predictions and to existing data, while further parameter-free predictions are discussed in this Symposium \cite{Mart}. It has been found \cite{proxy2,EPJA} that the proxy-SU(3) scheme leads to an explanation of the prolate over oblate dominance in deformed nuclei, determining in parallel  the border of the prolate to oblate transition. 

In the present work we discuss the origins of the particle-hole symmetry breaking which leads to the prolate 
over oblate dominance and argue that it is a consequence of the short range nature of the nucleon-nucleon interaction and the Pauli principle.

\bigskip\noindent
{\bf PARTICLE-HOLE SYMMETRY BREAKING IN NUCLEAR SHELLS} 
\medskip\noindent

The prolate over oblate dominance in deformed nuclei, discussed in the framework of proxy-SU(3) in Refs.
\cite{proxy2,EPJA}, is based on the breaking of the particle-hole symmetry within nuclear shells, as seen in Table I of Ref. \cite{proxy2}. In that table, the highest weight (h.w.) irreducible representations (irreps) 
$(\lambda,\mu)$, where $\lambda$ and $\mu$ are the Elliott quantum numbers \cite{Elliott1,Elliott2}, are shown, for each number of particles in each shell. In addition, the irreps having the highest eigenvalue 
of the second order Casimir operator of SU(3) \cite{Draayer}
\begin{equation}
C_2(\lambda,\mu) = (\lambda+\mu+3)(\lambda +\mu) -\lambda \mu, 
\end{equation}
 which do exhibit particle-hole symmetry within each shell, are shown. 

In the first half of each shell, the h.w.-irreps coincide with the irreps with the highest $C_2$ eigenvalue.
Differences appear as soon as the middle of the shell is crossed and they go on until the shell is almost filled, with only 4 holes remaining. It is interesting to study this difference and possibly locate its 
 origins. In doing so, it is useful to recall that  $C_2$ is known to be proportional to the 
 quadrupole-quadrupole interaction  \cite{Draayer}
\begin{equation}
C_2= {1\over 4} Q\cdot Q + {3\over 4} L^2,
\end{equation}
where $Q$ is the quadrupole operator and $L$ denotes the angular momentum.  The quadrupole-quadrupole interaction is known to the dominant in deformed nuclei \cite{Elliott1,Elliott2}. 

Since it is known that the h.w. irrep has to be a single one \cite{RBcode}, i.e., appearing only once in the relevant decomposition, we do not need to consider in detail all irreps for a given number of particles, as we did in Ref. \cite{SarSDANCA}, but only the single ones. 

Further simplification of the tables is possible by taking into account that in all algebras and for all even numbers of particles
the h.w. irreps possess even values of $\lambda$ and $\mu$. This is in agreement with the fact that the ground state of all even-even nuclei has zero angular momentum \cite{Casten}. $L=0$ states can occur only within $K=0$ irreps, where $K$ is the Elliott missing quantum number in the SU(3) to SO(3) decomposition, while $K=0$ irreps can occur only within even $\mu$ or $\lambda$ values \cite{Elliott1,Elliott2}.
As a consequence, we keep in the tables only the irreps with $\lambda$ even and $\mu$ even. 

Complete sets of irreps for even numbers of particles are given for U(6), U(10), U(15) and U(21) in Table I, while similar results for U(28) and U(36) are shown in Table II, respectively. One can make the following observations:

1) In all algebras the h.w. irreps in the first half of the relevant shell are the irreps possessing the maximum eigenvalue of the second order Casimir operator of SU(3), $C_2(\lambda,\mu)$. 

2) In all algebras particle-hole symmetry appears up to 4 particles. The irreps up to 4 particles 
in the first half of the shell are prolate, while the irreps for the last 4 particles in the second half of the shell are oblate. 

3) For a given algebra and for a given number of particles in the first half of the relevant shell, there are in general some prolate irreps of higher weight, followed by oblate irreps of lower weight. In this case, 
for the conjugate number of particles within the same shell (i.e., beyond the middle of the shell), the irrep appearing as the highest weight irrep is the conjugate of the oblate irrep having the highest eigenvalue of 
$C_2(\lambda,\mu)$ in the first half of the shell. In other words, while prolate irreps with the highest $C_2(\lambda,\mu)$ win in the first half of the shell, the conjugates of the oblate irreps of the first half of the shell with the highest $C_2(\lambda,\mu)$ become winners in the second half of the shell. In that case 
both winners, in the  first and in the second part of the shell, are prolate, a fact directly related to the prolate over oblate dominance.

4) However, there are cases, for small numbers of particles $\geq 4$, that no oblate irreps appear in the first half of the shell. In this case, the prolate irrep with the highest $C_2(\lambda,\mu)$ is the winner 
in the first half of the shell, while the conjugate of the prolate irrep with the  lowest 
$C_2(\lambda,\mu)$ in the first half of the shell (one could say ``the conjugate of the less prolate irrep'') becomes the winner in the second half of the shell. 
From Tables I and II one sees that 
this happens for 6 particles in U(10)-U(36), for 8 particles in U(21)-U(36), for 10 particles only in U(36). 
These are the only cases in which the h.w. irreps in the second part of the shell will be oblate. 
In other words, with the exception of the irreps related to the last 4 particles in each shell, mentioned in 2), oblate irreps appear only for 30-6=24 particles in U(15), for 42-6=36 and 42-8=34 particles in U(21),  
for 56-6=50 and 56-8=48 particles in U(28), for 72-6=66, 72-8=64, and for 72-10=62 particles in U(36). 
In other words, oblate irreps appear only just below the closing of the shell, a fact again directly related to the prolate over oblate dominance.

5) From the above it becomes clear that there is a particle-hole symmetry around the midshell, 
since the single irreps appearing in the second half of the shell are the conjugates of the irreps appearing 
in the first half of the shell. There is also some symmetry in the fact that in the first half of the shell
the prolate irreps with high $C_2(\lambda,\mu)$ are favored, while in the second half of the shell the formerly oblate irreps with high $C_2(\lambda,\mu)$ are favored. In the absence of formerly oblate irreps, the formerly prolate irreps with the lowest $C_2(\lambda,\mu)$ (the ``least prolate irreps'') are favored.    

\includegraphics[width=205mm]{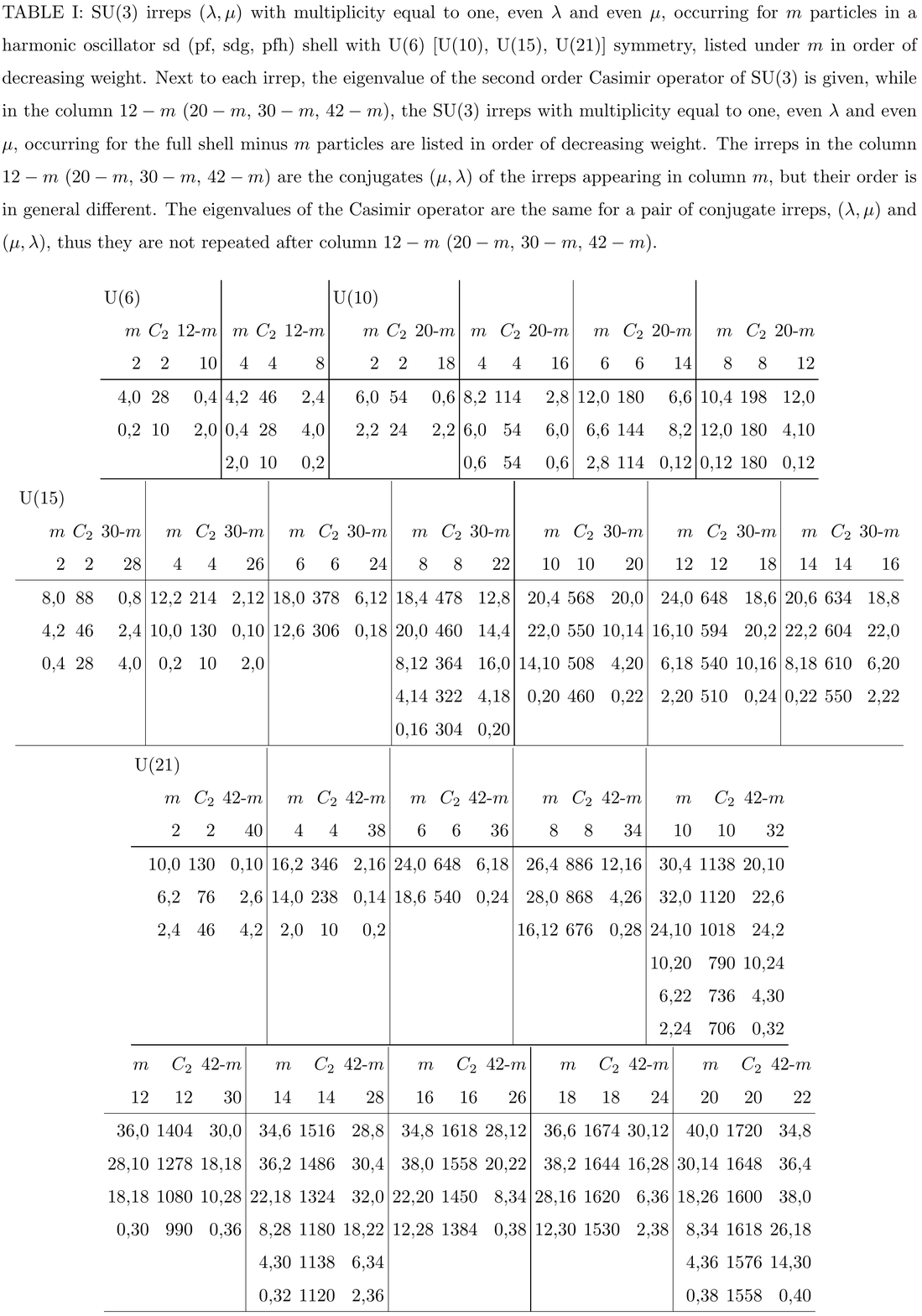}

\includegraphics[width=205mm]{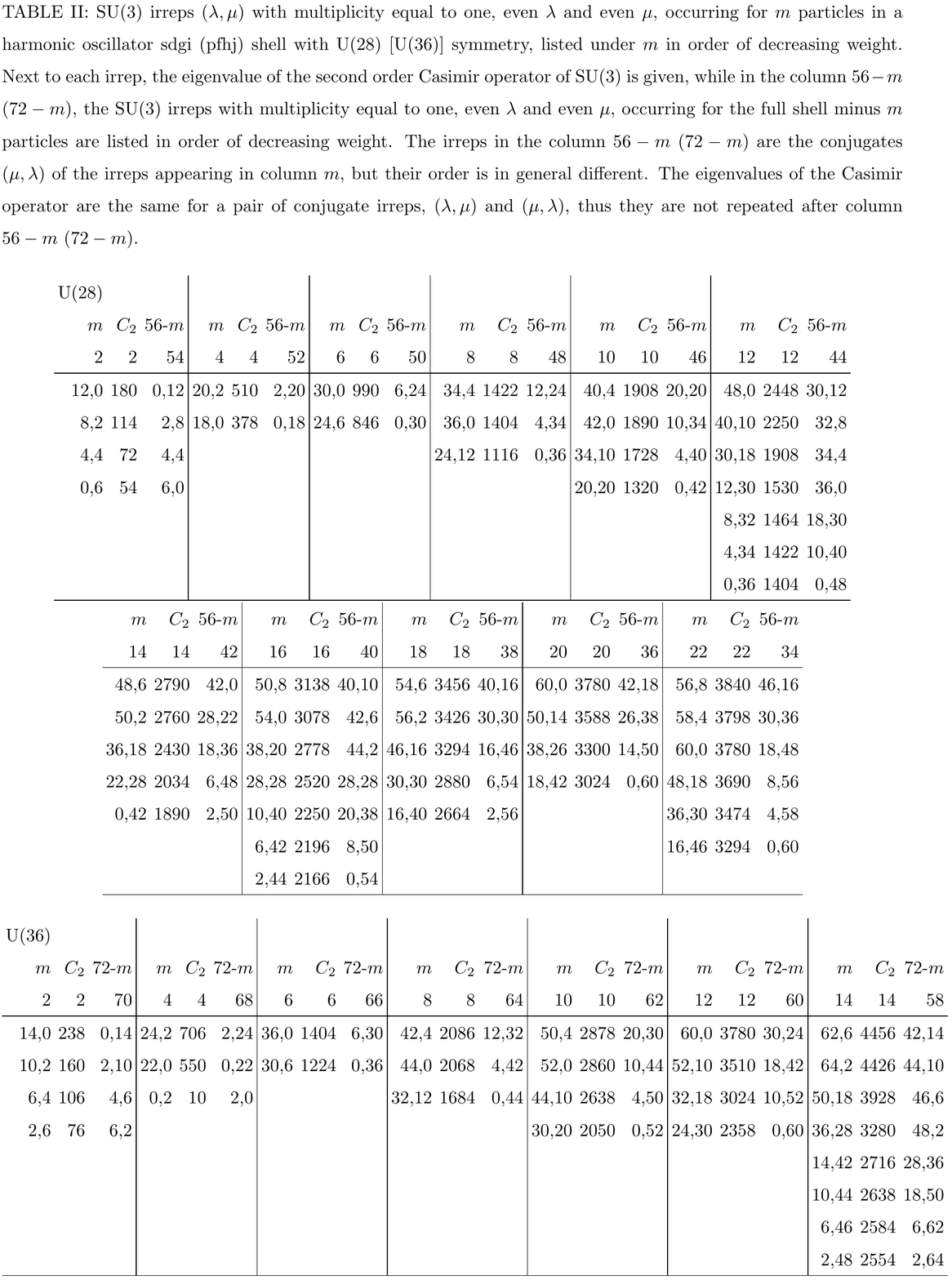}

\bigskip\noindent
{\bf THE ROLE OF THE SHORT RANGE INTERACTION} 
\medskip\noindent

The above observations become more transparent by taking into account the fact that the nucleon-nucleon
interaction is characterized by a short range \cite{Casten,Ring}, which favors maximal spatial overlaps occurring in the case of symmetrized spatial wave functions
\cite{Casten,Ring}. However, because of the Pauli principle, the spin-isospin part of the wave function has to be antisymmetric, in order to guarantee the antisymmetric character of the total wave function. 
It turns out that the highest weight irrep is the irrep with the maximum spatial symmetrization possible, given the restrictions imposed by the Pauli principle. 

This point can be clarified through an example. Consider the U(15) shell (sdg shell). If one has 10 particles in it, the h.w. irrep is (20,4). If one has 10 holes, i.e. 30-10=20 particles, the h.w. irrep is (20,0), while in the case of p-h symmetry one would have expected the conjugate irrep of (20,4), i.e. (4,20). One can plot the (20,0) and (4,20) irreps to see which is more symmetric. The Young diagram for (20,0)  consists of 20 boxes in the first row and no boxes in the second row. The Young diagram for (4,20) has 24 boxes in the first row and 20 boxes in the second row. In the Young diagrams, boxes in the same row mean symmetrization, boxes in the same column mean antisymmetrization. Therefore (20,0) is purely symmetric, while (4,20) contains lots of antisymmetrizations.  The most symmetric irrep gives the highest spatial overlaps, and it corresponds to the most antisymmetric spin-isospin irrep. This happens because the nucleon-nucleon interaction is of short range, thus it prefers the most symmetric spatial irreps. The (4,20) irrep does have a higher Casimir eigenvalue than the (20,0) irrep, but it is {\sl not} the one preferred by the short range interaction in combination with the Pauli principle.

One has to realize that the influence of the Pauli principle is not exhausted by imposing the antisymmetry of the spin-isospin part of the wave function. It does influence in parallel the structure of the spatial part of the wave function as well. There is no need for a specific Hamiltonian in order to see the above. Any nuclear Hamiltonian has to respect the Pauli principle and the short range nature of the nucleon-nucleon interaction. 

The fact that the quadrupole-quadrupole interaction is maximized in deformed nuclei in the first half of the shell is due to the Pauli principle and to the short range nature of the nucleon-nucleon interaction. In the second half of the shell it is not the irrep giving the highest quadrupole-quadrupole interaction (equivalently: the highest eigenvalue of the second order Casimir operator) the one which is preferred by the Pauli principle and the short range interactions. The Pauli principle is still there, thus the spin-isospin irrep still has to be most antisymmetric, thus the spatial irrep still  has to be most symmetric.

\bigskip\noindent
{\bf DISCUSSION}
\medskip\noindent

The main conclusion drawn is that the particle-hole symmetry breaking within a nuclear shell, 
the prolate over oblate dominance in deformed nuclei and the prolate-oblate transition come from the Pauli principle and the short range of the nucleon-nucleon interaction alone.

\bigskip\noindent
{\bf ACKNOWLEDGEMENTS}
\medskip\noindent

Work partly supported by the Bulgarian National Science Fund (BNSF) under Contract No. DFNI-E02/6, by the MSU-FRIB laboratory, by the Max Planck Partner group, TUBA-GEBIP, and by the Istanbul University Scientific Research Project No. 54135.

\end{document}